\documentclass{goslar}

\begin{document}

\title{Nonlinear von Neumann-type equations\footnote{
{\it New Insights in Quantum Mechanics\/}, H.-D. Doebner, S.T. Ali, M. Keyl,
and R.F. Werner, eds. (World Scientific, Singapore, 1999)}}

\author{Marek Czachor $^{1,2,3}$, Maciej Kuna $^{1,3}$, 
Sergiej B. Leble $^{1,4}$, \\
and Jan Naudts $^3$}

\address{$^1$ Wydzia{\l} Fizyki Technicznej i Matematyki Stosowanej\\
Politechnika Gda\'nska, 80-956 Gda\'nsk, Poland\\
$^2$ Arnold Sommerfeld Institute for Mathematical Physics\\
Technical University of Clausthal, 38678 Clausthal-Zellerfeld, Germany\\
$^3$ Departement Natuurkunde, Universiteit Antwerpen, UIA\\
2610 Antwerpen, Belgium\\
$^4$ Department of Theoretical Physics, Kaliningrad State University\\
236041 Kaliningrad, Russia}

\maketitle

\abstracts{We review some recent developments in the theory of nonlinear 
von Neumann equations. We distinguish between the von Neumann equation 
(which can be nonlinear) and the Liouville equation (which should be linear). 
Explicit examples illustrate 
the technique of binary Darboux integration of nonlinear density matrix 
equations and special attention is payed to the problem of how to find 
physically nontrivial `self-scattering' solutions.}

\section{Why density matrices?}

``It is a common error always to view a mixed state as describing a system 
that is actually in one of a number of different possible pure states, 
with specified probabilities. While this `ignorance interpretation' of the 
mixed state can indeed be a useful practical way to describe an ensemble 
of completely isolated systems, it entirely misses the deep and fundamental 
character of mixed states: If a system has any external correlations 
whatever, then its quantum state cannot be pure. Pure states are a rarity, 
enjoyed only by completely isolated systems. The states of externally 
correlated individual systems are fundamentally and irreducibly mixed.
This has nothing to do with `our ignorance'. It is a consequence of the 
existence of objective external correlation'' \cite{Mermin}. 

The quotation from the Mermin paper can serve as a motto for what we 
are going to present below. A physical system that is described by a state 
vector at time $t=0$ will remain in a pure state for all $t>0$  if and only 
if it will never interact with anything. An interaction leads to correlations,
and correlations mean non-product (entangled) states. A subsystem of a 
bigger composite 
system which, as a whole, is in an entangled state, is no longer described 
by a state vector but by a non-pure reduced density matrix. 
The density matrix is 
an entirely quantum object whose ontological status is the same as this of the 
state vector it originated from. A theory that deals only with pure states 
deals either with the entire Universe or 
with objects that cannot interact and are therefore unobservable. 

The above remarks are especially relevant if one thinks of nonlinear 
generalizations of quantum mechanics. `Nonlinear quantum mechanics' is 
traditionally associated with nonlinear Schr\"odinger equations. Such equations
have the general form
\bea
i|\dot\psi\rangle=H(\psi)|\psi\rangle.\label{1}
\eea
Assuming we have a concrete physical system which, when isolated, evolves 
according to (\ref{1}) we immediately face the question of how to describe 
this system during or after an interaction. The problem 
 has led to a great amount of confusion as to the real 
status of nonlinear generalizations of quantum mechanics. Obviously, the 
resulting misunderstandings involve an interpretation of experiments
testing linearity of quantum mechanics: A theory that does not tell us how to
describe external correlations does not produce testable predictions and, 
hence, cannot be tested. 

The problem of subsystems versus (\ref{1}) was, in our oppinion, solved
by Polchinski in \cite{P}. In short, his answer was the following: The 
equation describing a composite system is
\bea
i|\dot\Psi\rangle=\Big(
H_{\rm s}(\rho_{\rm s})
+
H_{\rm r}(\rho_{\rm r})
\Big)
|\Psi\rangle\label{2}
\eea
where the index `s' stands for a subsystem and `r' for the `rest' 
or `remaining systems'. The state vector $|\Psi\rangle$ represents the pure 
state of the composite `subsystem plus the rest' system,
and the $\rho$'s are the appropriate reduced density matrices. 
Let us note that, even if we assume, as we have done in (\ref{2}),  
that the composite system is in a pure state, the formalism inevitably leads 
us to density-matrix-dependent  nonlinear operators $H_{\dots}(\rho_{\dots})$.

To include interactions one can write 
\bea
i|\dot\Psi\rangle=\Big(
H_{\rm s}(\rho_{\rm s})
+
H_{\rm r}(\rho_{\rm r})
+
H_{\rm int}(\Psi)
\Big)
|\Psi\rangle\label{3}
\eea
with some interaction (linear or nonlinear) Hamiltonian operator 
$H_{\rm int}(\Psi)$. The main feature of (\ref{2}) and (\ref{3}) is the fact 
that for $H_{\rm int}(\Psi)=0$ (that is, when the interaction is over) 
the reduced density matrices satisfy the von Neumann-type equations
\bea
i\dot \rho_{\rm s}
&=&
[H_{\rm s}(\rho_{\rm s}),\rho_{\rm s}],\label{vNa}\\
i\dot \rho_{\rm r}
&=&
[H_{\rm r}(\rho_{\rm r}),\rho_{\rm r}].\label{vNb}
\eea
The form of these von Neumann equations illustrates the {\it locality\/} of 
the Polchinski formulation: Subsystems that do not interact do not `see' 
each other. It should be stressed that this property is typically  lost when
one considers different ways of extending the subsystem dynamics. 
An interesting discussion of the problem is given in these proceedings 
by L\"ucke \cite{L} who shows that there may exist local extensions different 
from those discussed by Polchinski. However, the nonlinear Schr\"odinger equations
they apply to are linearizable by a nonlinear gauge transformation 
\cite{G}. The Polchinski-type extensions bear some 
kind of universality but simultaneously are non-unique in the 
following sense: There exists an infinite number of inequivalent extensions 
which reduce to the same equation on product states. A more detailed 
discussion of this and related problems can be found in 
\cite{P,J,MC96,MC97,MCMK98}. 

From what we have written so far it follows that nonlinear von Neumann 
equations, whether we like it or not, will always appear when quantum 
nonlinearity occurs. A somewhat more radical viewpoint is suggested by the 
non-uniqueness of the Polchinski-type extensions. To illustrate the point 
consider a simple spin-1/2 nonlinear `average energy'
\bea
H(\psi)
&=&
\frac{\langle\psi|\sigma_z|\psi\rangle^2}{\langle\psi|\psi\rangle}.
\label{sigma}
\eea
where $\sigma_z$ is the Pauli matrix. 
This type of nonlinearity was considered in experiments involving 2-level
atoms \cite{MC96}. 
Now, how to write $H(\rho)$ on the basis of $H(\psi)$? A natural guess is 
\bea
H(\rho)
&=&
\frac{({\rm Tr\,}\sigma_z\rho)^2}{{\rm Tr\,}\rho}.
\eea
but 
\bea
H(\rho)
&=&
\frac{{\rm Tr\,}\sigma_z\rho\sigma_z\rho}{{\rm Tr\,}\rho}.
\eea
and an infinite number of similar fuctions would do as well. 
They all reduce to (\ref{sigma}) if $\rho=|\psi\rangle\langle\psi|$. 
However, if we {\it start\/} with some $H(\rho)$ then the whole ambiguity 
disappears. This suggests that looking for a fundamental level of nonlinear
quantum dynamics one should begin with von Neumann and not Schr\"odinger 
equations. 

Linear Schr\"odinger equation, as seen from a classical perspective, is 
simply
the equation of motion of
 an infinite-dimensional Hamiltonian system with average energy 
playing a role of Hamiltonian function. For a detailed presentation 
of the formalism see the paper by Cirelli {\it et al.\/} 
published in these proceedings \cite{C}. 
Linear von Neumann equation is also
the equation of motion of
a classical Hamiltonian system with 
average energy in the role of the Hamiltonian function. Geometrically, 
density matrices form an infinite-dimensional Poisson manifold with  
$gl(\infty)$ Lie-Poisson bracket. This is described in the present volume 
by B\'ona \cite{B} and for this reason we will not spend much 
time on details of the Lie-Poisson formulation. 

\section{Nonlinear von Neumann equation}

Let us denote by $\rho_a$ a matrix element of the density matrix $\rho$ taken 
in some basis. The Lie-Poisson version of the {\rm linear\/} 
von Neumann equation is 
\bea
i\dot\rho_a=\{\rho_a,\langle H\rangle\}\label{LP}
\eea
where $\langle H\rangle={\rm Tr\,}H\rho$ is the Hamiltonian function. 
Nonlinear quantum mechanics {\it \'a la\/} B\'ona and Jordan \cite{J,B} is 
based on the same Lie-Poisson structure but with $\langle H\rangle$ replaced 
by a nonlinear Hamiltonian function $H_{\rm gen}$. 
The dynamics given by (\ref{LP}) or its nonlinear generalization 
\bea
i\dot\rho_a=\{\rho_a,H_{\rm gen}\}\label{nLP}
\eea
defines a Hamiltonian flow on the Poisson 
manifold of states. Having such a flow one can consider a classical probability
density $w(\rho)$ and its associated Liouville equation 
\bea
i\dot w=\{w,H_{\rm gen}\}\label{L}.
\eea
The classical probability density has a clear physical meaning: It describes 
a classical lack of knowledge about the state of a quantum source. In any 
experiment one faces this type of clasical uncertainty and all experimental 
averages one measures in a lab are of the form
\bea
\langle A\rangle_{\rm exp}
=
\int d\rho \,w(\rho){\rm Tr\,}A\rho\label{Aexp}
\eea
with $\int d\rho$ meaning an integration over the parameters
defining initial conditions for the quantum dynamics. The Liouville equation
(\ref{L}) is {\it linear\/} in $w$ independently of whether the von Neumann 
equation (\ref{nLP}) is linear in $\rho_a$. 
Moreover, the probability density $w$
is directly accessible to the experimentalist and reflects the classical
configuration of the experimental setting. For this reason it is very 
important that the Liouville equation is linear. Formula (\ref{Aexp}) 
shows that it is practical to introduce the object 
\bea
\rho_{\rm exp}
=
\int d\rho \,w(\rho)\rho\label{rhoexp}
\eea
satisfying 
\bea
\langle A\rangle_{\rm exp}
=
{\rm Tr\,}A\rho_{\rm exp}
\eea
and playing a role of a {\it semiclassical\/} density matrix. 
The `common error', mentioned by Mermin in the quotation we have started 
with, is the belief that all density matrices one encounters in quantum 
mechanics have such a semiclassical origin. The `truely quantum' 
density matrices one obtains by reduction of entangled states to subsystems 
can be written in different bases of `pure states' and all such bases 
are regarded as physically equivalent. Putting it differently, no particular
decomposition of such a $\rho$ into a convex combination of projectors is 
physically special. This is one of the important impossibility principles 
of quantum mechanics \cite{M}. On the other hand, the decomposition defined
by $w$ is not only very special but, actually, is even uniquely given by 
the form of the experiment. An experimentalist can arbitrarily tamper with 
$w$ but different convex combinations forming a concrete $\rho$ are definitely
out of his reach. 

One can imagine also physical situations which are somehow in-between the cases
we have just described. For example, consider an entangled 
pair of particles and an 
experiment where we have a shutter which is opened whenever a particle
labeled `1', say, is measured and a concrete result (say, spin `up' or 
`down' in a `$z$' direction) is
found. Each time the shutter is opened the particle labeled `2' leaves 
a box and, hence, the box is a source of particles in a concrete mixed state 
which depends on the entangled state of the pair. The mixed state, 
as depending on the macroscopic and clasically controlled actions undertaken
by the experimentalist, is no longer `fully quantum'. The resulting mixture 
is of a $\rho_{\rm exp}$ type and there is no reason for the corresponding
dynamics of the density matrix to be nonlinear. It seems reasonable to
assume that nonlinear quantum dynamics of mixed states can occur only in cases 
where the very form of the `pure-state' decomposition is {\it in 
principle\/} out of control. Such an impossibility principle seems to 
eliminate all the problems analyzed by Polchinski in \cite{P}. 

The distinction between the von Neumann equation (\ref{nLP}) and the Liouville
equation (\ref{L}) is therefore essential. One should not use 
the misleading term the `Liouville-von Neumann equation' which suggests that 
the von Neumann equation is simply a quantized version of the linear Liouville 
equation and, accordingly, {\it must\/} be linear as well. 

Historically, the linearity of the standard von Neumann equation seems to 
have its roots in the linearity of the Schr\"odinger equation. 
The two equations, 
\bea
i|\dot\psi_k\rangle=H|\psi_k\rangle, 
\quad
-i\langle\dot \psi_k|
=
\langle\psi_k|H
\eea
combined with $\rho=|\psi_k\rangle\langle\psi_k|$ imply
\bea
i\dot \rho
&=&
[H,\rho].\label{lvN}
\eea
Since the equation is linear, one can consider more general convex 
combinations 
\bea
\rho=\sum_k \lambda_k |\psi_k\rangle\langle\psi_k|
\eea
which again satisfy (\ref{lvN}). This simple argument looks so natural that 
one may have not even noticed the additional assumption we have smuggled in. 
Indeed, we have started with  (\ref{lvN}) which was valid for pure states, 
that is those satisfying $\rho^2=\rho$, and by linearity extended the argument 
to $\rho\neq \rho^2$. But a pure $\rho$ has eigenvalues 0 and 1. Therefore 
any function $f$ satisfying $f(0)=0$ and $f(1)=1$ will satisfy 
$f(\rho)=\rho$ if $\rho$ is pure! It follows that the linearity of the 
Schr\"odinger equation implies at most an equation of the form 
\bea
i\dot \rho
&=&
[H,f(\rho)]\label{fvN}
\eea
with $[H,f(\rho)]=[H,\rho]$ for $\rho^2=\rho$. In the next sections we shall 
devote much attention to the so-called Euler-Arnold-von Neumann equation
obtained if $f(x)=x^2$. 

Now, what is the relation
between (\ref{fvN}) and (\ref{nLP})? The answer is rather surprising: 
(\ref{fvN}) is an example of (\ref{nLP}) with an appropriate choice of 
$H_{\rm gen}$ \cite{MCJN}. 
To see how this works assume that for $x\in [0,1]$ the function
$f$ has a convergent Taylor expansion 
\bea
f(x)=\sum_{k=0}^\infty f_k(x-a)^k \label{f}
\eea
and define 
\bea
H_{\rm gen}(\rho)
&=&
{\rm Tr\,}\big(f(\rho)H\big)
\eea
(in the context of \cite{MCJN} it is important that $H_{\rm gen}(\rho)$ is 
1-homogeneous; this complication is left out here). 
Then (\ref{nLP}) is equivalent to 
\bea
i\dot \rho=[\hat H(\rho),\rho]\label{effvN}
\eea
where the effective nonlinear Hamiltonian is defined by the functional 
derivative \cite{MCJN}
\bea
\hat H(\rho)
=
\frac{\delta}{\delta\rho}H_{\rm gen}(\rho)
=
\sum_{k=1}^\infty f_k\sum_{n=0}^{k-1}(\rho- a{{\bf 1}})^{k-1-n}
H
(\rho- a{{\bf 1}})^{n}
\eea 
and 
\bea
i\dot \rho=[\hat H(\rho),\rho]=[H,f(\rho)]. 
\eea
The form (\ref{effvN}) implies that $C_n=
{\rm Tr\,} (\rho^n)$ is time-independent for any natural $n$ (such $C_n$ are
Casimir functions for $\{\cdot,\cdot\}$) from which it follows that the 
spectrum of a solution $\rho$ of (\ref{fvN}) 
 has to be time-invariant \cite{MCMM}. As we can see, although the dynamics 
of $\rho$ is nonlinear for non-pure $\rho$, we are nevertheless still quite
close to linear quantum mechanics. Looking at the explicit solutions for 
$f(x)=x^2$ we will see that we are, in fact, {\it surprisingly\/} 
close to the linear dynamics but with very specific and subtle new nonlinear
effects at hand. 

It is quite clear there are fundamental reasons for investigating nonlinear 
density matrix equations. But one does not have to believe in nonlinear 
quantum mechanics to investigate nonlinear evolutions of mixed states. 
It is known that nonlinear von Neumann equations arise very naturally 
in various mean-field approaches \cite{R} where they are solved either 
numerically or by approximate methods. The program we are developing 
aims at finding {\it exact\/} techniques of solving 
nonlinear von Neumann equations.

\section{Lax representation of von Neumann equations and binary Darboux 
covariance}

The class of equations we have under some control is of the form
\cite{SLMC,MKMCSL}
\bea
i\dot\rho
&=&
[\hat H(\rho),\rho]
=
{\sum_{k=0}^n} [A^{n-k}\rho A^k,\rho]
={\sum_{k=0}^n} [A^{n-k},\rho A^k \rho] \label{n-eq}
\eea
where $A$ is a time-independent self-adjoint operator. 
For $n=1$ and $A=H$
one finds 
\bea
\hat H(\rho)=H \rho+ \rho H
\eea
and the equation becomes the Euler-Arnold-von Neumann equation
\bea
i\dot\rho
=[H \rho+ \rho H,\rho]=[H,\rho^2].
\eea
For all $n$ the von Neumann equations involve Hamiltonians $\hat H(\rho)$ 
which are {\it linear\/} in $\rho$, and therefore the nonlinearities 
of the equations are quadratic. This property 
is important as it seems that only the 
quadratic nonlinearities can be treated by the binary Darboux transformation.
In fact, what  makes the class so interesting is the knowledge
of its binary-Darboux-covariant Lax representation
\bea
z_\mu |\varphi\rangle 
&=& 
(\rho-\mu A)|\varphi\rangle,\label{1a}\\
i|\dot\varphi\rangle 
&=&
\Big(
\sum_{k=0}^n A^{n-k}\rho A^k-\mu A^{n+1}\Big)|\varphi\rangle,
\label{1b}
\eea
where $z_\mu$, $\mu\in{\bf C}$. The necessary condition for the 
solution $|\varphi\rangle$ of 
(\ref{1a})-(\ref{1b}) to exist is that the relation (\ref{n-eq}) 
between $A$ and $\rho$ holds true. 
The idea of the Darboux transformation is, given
a density matrix $\rho$ and a solution $|\phi\rangle$ of 
(\ref{1a})-(\ref{1b}), to find 
a new solution $|\varphi[1]\rangle$ and a new $\rho[1]$ which are again 
related by the pair (\ref{1a})-(\ref{1b}) with $A$ unchanged. 
The fact that $|\varphi[1]\rangle$ is a solution means that the nesessary 
condition for its existence is satisfied, and this implies that $\rho[1]$
and $A$ are again related by (\ref{n-eq}). 
One may think of $\rho$ and $\rho[1]$ as `ground' and `first excited'
states, and the Darboux transformation is a `creation operator'. 

The so-called binary Darboux transformation is constructed as follows (for more
details and generalizations cf. \cite{SLMC,MKMCSL,ZL,Leble,U}).
Take some linear operators $V$ and $J$, a parameter $s$, and three
complex numbers $\mu$, $\nu$, and $\lambda$. Now consider three linear
equations 
\bea
i\partial_s|\varphi\rangle 
&=&
\big(V-\mu J)|\varphi\rangle, \label{ket}\\
-i\partial_s\langle\chi|
&=&
\langle\chi|\big(V-\nu J), \label{bra1}\\
-i\partial_s\langle\psi|
&=&
\langle\psi|\big(V-\lambda J). \label{bra2}
\eea
It is important that (\ref{bra1}) and (\ref{bra2}) involve bras and 
(\ref{ket}) involves a ket. Define
\bea 
P&=&\frac{|\varphi\rangle\langle\chi|}{\langle\chi|\varphi\rangle}\label{P1}\\
V[1] 
&=& V + (\mu-\nu)[P,J]\label{VBDT}\\
\langle\psi[1]|
&=&
\langle\psi|\Big({{\bf 1}}
- \frac{\nu-\mu}{\lambda-\mu}P\Big)\label{BDT}
\eea
(the fact that we transform here $\langle\psi|$ and not $|\varphi\rangle$ 
or $\langle\chi|$ is not essential; one can construct similar transformations
of any of these solutions with the help of the remaining two).
A straightforward calculation then shows that 
\bea
-i\partial_s\langle\psi[1]|
&=&
\langle\psi[1]|\big(V[1]-\lambda J).\label{eq[1]}
\eea
To apply the above technique to our von Neumann equations we take three 
{\it pairs\/}: (\ref{1a})-(\ref{1b}), which we already have, and two more 
 with parameters $\lambda$, $z_\lambda$,
$\nu$, $z_\nu$:
\bea
z_\lambda \langle\psi| 
&=& 
\langle\psi|(\rho-\lambda A),\label{2a}\\
-i\langle\dot\psi| 
&=&
\langle\psi|\Big(
\sum_{k=0}^n A^{n-k}\rho A^k-\lambda A^{n+1}\Big),
\label{2b}\\
z_\nu \langle\chi| 
&=& 
\langle\chi|(\rho-\nu A),\label{3a}\\
-i\langle\dot\chi| 
&=&
\langle\chi|\Big(
\sum_{k=0}^n A^{n-k}\rho A^k-\nu A^{n+1}\Big).\label{3b}
\eea
In what follows the $\mu$- and $\nu$-pairs (\ref{1a})--(\ref{1b}) 
and (\ref{3a})--(\ref{3b}) will be used to
define the binary transformation of the conjugated
$\lambda$-pair (\ref{2a})--(\ref{2b}).

Defining $P$ as above and 
\bea
\rho[1]=\rho+(\mu-\nu)[P,A]\label{rho[1]}
\eea
we find \cite{MKMCSL}
\bea
z_\lambda \langle\psi[1]|
&=& 
\langle\psi[1]|(\rho[1]-\lambda A)\\
-i\langle\dot\psi[1]| 
&=&
\langle\psi[1]|\Big(
\sum_{k=0}^n A^{n-k}\rho[1] A^k-\lambda A^{n+1}\Big). 
\eea
If all the objects necessary for the construction of $\langle\psi[1]|$ exist,
then the necessary condition for its existence must be fulfilled, and this
means
\bea
i\dot\rho[1]
&=&
{\sum_{k=0}^n} \big[A^{n-k}\rho[1] A^k,\rho[1]\big]
={\sum_{k=0}^n} \big[A^{n-k},\rho[1] A^k \rho[1]\big] \label{n-eq[1]}.
\eea
Having one solution $\rho$ we have managed to produce another 
solution $\rho[1]$. In this respect the Darboux transformation is a really 
wonderful device. However, from time to time surprises can occur. 
It is clear that if all the assumptions we have made are satisfied then 
$\rho[1]$ must be a solution. But the general theorem does not guarantee 
that the solution is nontrivial! In fact $\rho=A$, $\rho={{\bf 1}}$, or even 
$\rho=0$
are also solutions  of (\ref{n-eq}), so can we guarantee that such pathological
cases are excluded and $\rho[1]$ is physically interesting? 

The answer depends on what is meant by `interesting'. 
In general, each Darboux transformation has an inverse. In classical 
problems, such as the Korteweg-de Vries equation, it is typical to start 
with a trivial solution $u=0$ and $u[1]$ is already a soliton, 
definitely a highly nontrivial solution \cite{MS}.
This means there exists a transformation that maps a soliton into 0. 

We will now show that this type of pathology is excluded if a binary 
transformation of the type we use is considered. What is not excluded, 
however, 
are the cases when, say, $\rho[1]=\rho$ or $\rho[1]=T\rho T^{-1}$, with $T$
a time-independent unitary transformation. Such a possibility leads to 
nontrivial practical complications. The next theorem is of fundamental 
importance and its proof is so elementary that we can give it here 
\cite{MKMCSL}.
Consider the following three general Lax pairs 
\bea
z_\mu |\varphi\rangle 
&=& 
(\rho-\mu A)|\varphi\rangle ,\label{L1a}\\
i|\dot \varphi\rangle 
&=&
\big(V(\rho) - \mu J\big)|\varphi\rangle,\label{L1b}\\
z_\nu \langle\chi| 
&=& 
\langle\chi|(\rho-\nu A), \label{L3a}\\
-i\langle\dot \chi|
&=&
\langle\chi| \big(V(\rho) - \nu J\big),\label{L3b}\\
z_\lambda \langle\psi|
&=& 
\langle\psi|(\rho-\lambda A),\label{L2a}\\
-i\langle\dot \psi|
&=&
\langle\psi| \big(V(\rho) - \lambda J\big).\label{L2b}
\eea
The only assumption we make about $\rho$, $J$, $A$, and 
$V(\rho)$ is the covariance of
(\ref{L2a})--(\ref{L2b}) under the binary Darboux transformation
constructed with the help of $\langle\chi|$ and $|\varphi\rangle $. 

\medskip
\noindent
{\bf Theorem~1.} Under the above assumptions the binary Darboux
transformation $\rho\mapsto\rho[1]$ is a similarity
transformation,
$\rho[1]=T\rho T^{-1}$, where
\bea
T &=& {{\bf 1}}+\frac{\mu-\nu}{\nu}P=e^{P\ln\frac{\mu}{\nu}}
\eea
{\it Proof\/}: 
By definition $\rho[1]=\rho+(\mu-\nu)[P,A]$. 
Eqs.~(\ref{L1a}), (\ref{L3a}) imply 
\bea
z_\mu P
&=& 
(\rho-\mu A)P\label{LP1a}\\
z_\nu P
&=& 
P(\rho-\nu A)\label{LP3a}
\eea
from which it follows that 
\bea
P(\rho-\mu A)P
&=& 
(\rho-\mu A)P\label{PLP1a}\\
P(\rho-\nu A)P
&=& 
P(\rho-\nu A).\label{PLP3a}
\eea
Multiplying (\ref{PLP1a}) by $\nu$, (\ref{PLP3a}) by $\mu$, and
subtracting the resulting equations we get 
\bea
[P,A]=\frac{\nu-\mu}{\mu\nu} P\rho P-\frac{1}{\mu} \rho P+\frac{1}{\nu}
P\rho. 
\eea
Inserting this expression into the definition of $\rho[1]$ we
obtain 
\bea
\rho[1]=
\Big({{\bf 1}}+\frac{\mu-\nu}{\nu}P\Big)
\rho
\Big({{\bf 1}}+\frac{\nu-\mu}{\mu}P\Big).\label{central}
\eea
Q.E.D.

The theorem explains why $\rho[1]=0$ etc. is excluded if $\rho$ is a density 
matrix: Spectra of $\rho$ and $\rho[1]$ must be identical. 
This property is characteristic of all equations that are compatibility 
conditions for binary-Darboux-covariant Lax pairs. The result will, in 
general, not hold if one considers a  binary-Darboux-covariant 
zero-curvature pair of the type used for example in \cite{U}. 

The next important result, whose proof follows from a straightforward 
calculation, is the covariance under `spectrum shifting and rescaling'.

\medskip
\noindent
{\bf Theorem~2.} 
Assume $[X,A]=[X,\rho]=0$, $Y\in {\bf R}$, and $\rho$ is a solution of 
(\ref{n-eq}). Then 
\bea
\rho_X(t)
&=&
e^{-i(n+1)XA^nt}\big(\rho(t)+X\big)e^{i(n+1)XA^nt}\\
\rho_Y(t)
&=&
Y\rho(Yt)
\eea
also satisfy (\ref{n-eq}).

\section{Two strategies leading to `interesting' solutions}

The problem with the Darboux transformation is that in order to find 
a solution $\rho[1]$ we must already {\it know\/} somehow another solution 
$\rho$ to start with. As we have seen above, there are some obvious 
solutions such as 0, but by Theorem~1 they will not lead to anything 
nontrivial. Still, the case is not completely hopeless. 

\subsection{The first strategy}

The strategy works for the Euler-Arnold-von Neumann equation \cite{SLMC}
\bea
i\dot \rho
&=&
[H,\rho^2].\label{EA}
\eea
In all the examples discussed below we assume $\nu=\bar \mu$ and  
$|\varphi\rangle=|\chi\rangle$ since this guarantees that hermiticity
is conserved by the transformation.
A pure state $\rho=\rho^2$ is simultaneously a solution 
of (\ref{EA}) and of the linear von Neumann equation with the same $H$. 
Therefore, pure state 
solutions of the linear equation form a nontrivial subset of solutions 
of the nonlinear one. Unfortunately, by Theorem~1 we will have 
$\rho[1]=\rho[1]^2$ and although such $\rho[1]$ cannot be claimed `trivial'
they are nevertheless quite `uninteresting'. 
Interesting solutions are obtained if one starts with 
$\rho$ satisfying $[H,\rho^2-a\rho]=0$, for some $a\in {\bf R}$, but such 
that the operator $\Delta_a:= \rho^2-a \rho\neq 0$ is not a constant times 
${\bf 1}$.
In this case \cite{SLMC}
\bea
\rho(t)=e^{-ia Ht}\rho(0)e^{ia Ht}
\eea
and 
\bea
\rho[1](t)
&=&
e^{-iaHt}\Big(\rho(0)+(\mu-\bar\mu)F_a(t)^{-1}
 e^{-\frac{i}{\mu}\Delta_a t}
\big[|\varphi(0)\rangle\langle\varphi(0)|,H\big]
e^{\frac{i}{\bar \mu}\Delta_a t}\Big)e^{iaHt},\nonumber\\
&=:&
e^{-ia Ht}\rho_{\rm int}[1](t)e^{ia Ht}\label{rho[1]'}
\eea
where $\mu$ is a
complex parameter of the Darboux transformation,
$|\varphi(0)\rangle$ a solution of the Lax pair at $t=0$, and
$$F_a(t)=
\langle\varphi(0)|
\exp\Big(
i
\frac{\mu-\bar\mu}{|\mu|^2}\Delta_a t
\Big)|\varphi(0)\rangle.$$ 
After this first step we know were to look, but we still have to find
an appropriate $\rho$ and $|\varphi\rangle$. It turns out we need three 
more tricks which are best illustrated by the following example 
(see the Appendix).

Consider the Hamiltonian
\bea 
H=
\left(
\begin{array}{ccc}
0 & 1 & 0\\
1 & 0 & 0\\
0 & 0 & \frac{1}{\sqrt{2}}
\end{array}
\right)\label{H}
\eea
and take $\mu=i$ (for real $\mu$ the binary transformation is
trivial). We have to find an appropriate initial condition $\rho(0)$. 
The first trick we have mentioned is to begin with a solution which is 
neither normalized nor positive.
Such a non-density-matrix solution will be denoted by $\xi $
instead of $\rho$. 
We can always make it 
positive and normalized by using the transformations of Theorem~2. 

We take $\xi (t)=e^{-iHt}\xi (0)e^{iHt}$ with 
\bea 
\xi (0)=
\left(
\begin{array}{ccc}
\frac{1}{2}+ \frac{\sqrt{2}}{2} & 0  & 0\\
0 & \frac{1}{2}- \frac{\sqrt{2}}{2}  & 0\\
0 & 0 & \frac{1}{2}
\end{array}
\right).
\eea
The first two matrix elements on the diagonal are the solutions of the 
equation $x^2-x=1/4$ (this is the second trick) and for this reason
\bea 
\Delta_1=
\xi (0)^2-\xi (0)=\xi (t)^2-\xi (t)=
\frac{1}{4}
\left(
\begin{array}{ccc}
1 & 0  & 0\\
0 & 1  & 0\\
0 & 0 & -1
\end{array}
\right)
\eea
which obviously commutes with $H$ ($1/4$ is not essential here; this could be
basically any nonzero number). Therefore $[H,\xi (t)^2]=[H,\xi (t)]$ even though
$\xi (t)^2\neq \xi (t)$ and $[H,\xi (t)]\neq 0$. It may happen that appropriate 
solutions of the quadratic equation $x^2-x=x_0$ do not exist. But to make the 
trick work it is sufficient to find solutions of $x^2-ax=x_0$ with some $a$
and then we will have $\Delta_a$ instead of $\Delta_1$. The third trick is to
choose $\xi (0)$ and $|\varphi\rangle$ in such a way that the solution of the 
Lax pair is not an 
eigenstate of $\Delta_a$ (since the contributions from $F_a(t)$,
$\exp(-\frac{i}{\mu}\Delta_a t)$, and $\exp(\frac{i}{\bar \mu}\Delta_a t)$
would cancel one another in (\ref{rho[1]'}), and the internal part of 
(\ref{rho[1]'}) would
become time-independent --- that is exactly what we want to avoid).

The eigenvalues of $\xi (0)-iH$ are $z_\pm=(1\pm i\sqrt{2})/2$
and $z_-$ has degeneracy 2. The two eigenvectors
corresponding to $z_-$ are orthogonal:
\bea
|\varphi_1\rangle
=
\left(
\begin{array}{c}
0\\
0\\
1
\end{array}
\right),
\quad
|\varphi_2\rangle
=
\frac{1}{\sqrt{2}}
\left(
\begin{array}{c}
e^{i\pi/4}\\
1\\
0
\end{array}
\right).
\eea
Taking an arbitrary linear combination of them, say, 
\bea
|\varphi(0)\rangle=
\frac{1}{\sqrt{2}}\Big(|\varphi_1\rangle+|\varphi_2\rangle\Big)
\eea
we obtain an eigenvector all the three components of which are non-vanishing. 
This implies that the unitary transformation $T$ occuring in Theorem~1 will 
not be block-diagonal.  The fact that the two 
orthogonal eigenvectors correspond to the same eigenvalue of $\xi -\mu H$ 
and to different eigenvalues of $\Delta_a$  
allows us to construct a nonlinear dynamics involving 
the entire 3-dimensional space and automatically guarantees that 
$|\varphi\rangle$ is not an eigenstate of $\Delta_a$.  We find 
\bea
\xi [1](t)
&=&
e^{-iHt}\xi _{\rm int}[1](t)
e^{iHt}\label{int}
\eea
with
\bea
\xi _{\rm int}[1](t)
=
\left(
\begin{array}{ccc}
\frac{1 + \sqrt{2}}{2} - \frac{\sqrt{2}}{1 + e^t} & 0 &
   \frac{-1 - i}{2\sqrt{2}\cosh(t/2)}\\
0 & \frac{1 - \sqrt{2}}{2} + \frac{\sqrt{2}}{1 + e^t} & 
\frac{1}{2\cosh(t/2)}\\ 
\frac{-1 + i}{2\sqrt{2}\cosh(t/2)} & \frac{1}{2\cosh(t/2)} & \frac{1}{2}
\end{array}
\right).
\label{int'}
\eea
$\xi (t)$ is not yet a density matrix since it has a negative eigenvalue 
$(1-\sqrt{2})/2$ and its trace is $3/2$. Now it is time to use transformations
from Theorem~2. Set $X=\frac{\sqrt{2}-1}{2}{\bf 1}$ and $Y=\sqrt{2}/3$. 
A combination of the two transformations leads to the final solution
\bea
\rho_{XY}[1](t)
&=&
e^{-i\frac{2}{3}Ht}\rho_{\rm int}(t)e^{i\frac{2}{3}Ht}
\label {finsol}
\eea
where 
\bea
\rho_{\rm int}(t)
=
\frac{\sqrt{2}}{3}
\left(
\begin{array}{ccc}
\sqrt{2} - \frac{\sqrt{2}}{1 + e^{\sqrt{2}t/3}} & 0 &
   \frac{-1 - i}{2\sqrt{2}\cosh[t/(3\sqrt{2})]}\\
0 & \frac{\sqrt{2}}{1 + e^{\sqrt{2}t/3}} & 
\frac{1}{2\cosh[t/(3\sqrt{2})]}\\ 
\frac{-1 + i}{2\sqrt{2}\cosh[t/(3\sqrt{2})]} & \frac{1}{2\cosh[t/(3\sqrt{2})]}
 & \frac{\sqrt{2}}{2}
\end{array}
\right).
\eea
The density matrix $\rho_{XY}[1](t)$ has eigenvalues $2/3$, $1/3$, and 0.
For $|t|\to\infty$ one gets the dynamics asymptotically {\it linear\/}
but with different asymptotics for negative and positive times, and a kind of 
`self-scattering' (or `phase transition' as we called it in \cite{SLMC}) 
around $t=0$. To have some qualitative feel of what 
happens consider averages of spin-1 matrices
\bea
J_x
=
\left(
\begin{array}{ccc}
0 & 0 & 0\\
0 & 0 & i\\
0 & -i & 0
\end{array}
\right),
\quad
J_y
=
\left(
\begin{array}{ccc}
0 & 0 & -i\\
0 & 0 & 0\\
i & 0 & 0
\end{array}
\right),
\quad
J_z
=
\left(
\begin{array}{ccc}
0 & i & 0\\
-i & 0 & 0\\
0 & 0 & 0
\end{array}
\right).\nonumber
\eea
The figures illustrate the effect. 
Figures 1 and 2 show the evolution of the $x$--$y$ projection  
of $\langle {\bf J}\rangle$. The magnitude of the oscillation goes very 
quickly to 0 as $|t|$ grows. Figure~3 shows the dynamics of 
the $z$ component of  $\langle {\bf J}\rangle$. 
The process of `self-scattering' maps one asymptotically linear solution 
into another.  
\begin{figure}
\epsfig{figure=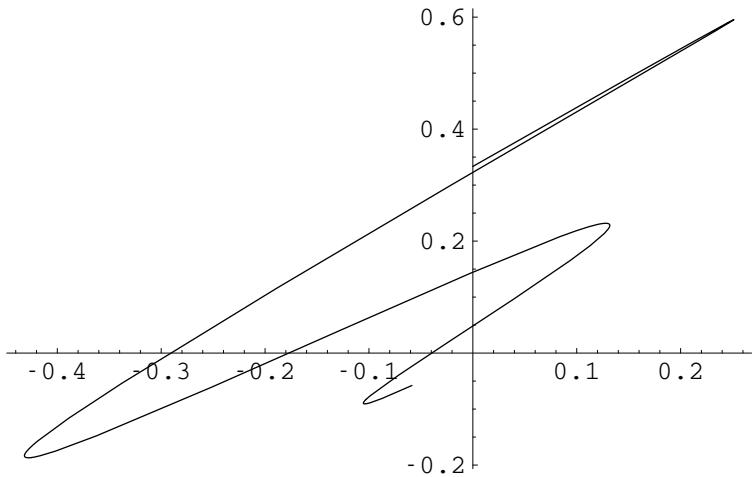}
\caption{Projection of the average spin on the $x$--$y$ plane  
for times $0\leq t\leq 10$ (in dimensionless units). The amplitude of 
the oscillation decreases with time.}
\end{figure}
\begin{figure}
\epsfig{figure=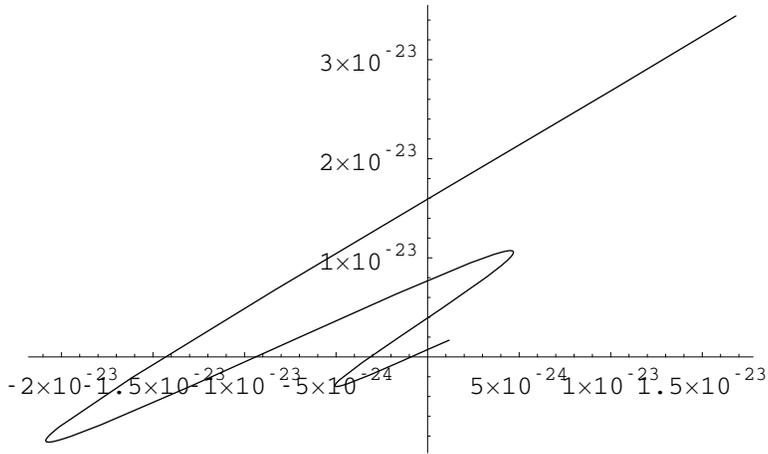}
\caption{Seeds of destruction of the past asymptotic state: 
The same dynamics as in Fig.~1 but for 
times $-230\leq t\leq -220$. The amplitude is $10^{22}$ times 
smaller than at Fig.~1. The amplitude of 
the oscillation increases with time.}
\end{figure}   
\begin{figure}
\epsfig{figure=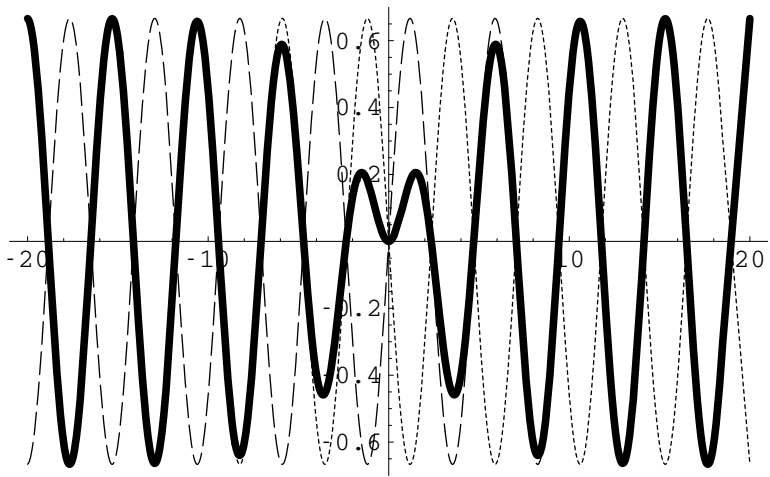}
\caption{The `self-scattering': Average $\langle J_z\rangle$ as a 
function of time (solid). 
The dynamics is asymptotically linear but with different 
asymptotics for $t\to +\infty$ (dashed) and $t\to -\infty$ (dotted).
The self-scattering phase shift is clearly visible.}
\end{figure}
A discussion of more realistic systems, including a one-dimensional 
harmonic oscillator, can be found in \cite{SLMC}.  

\subsection{The second strategy}

The first strategy leads to solutions involving a finite number of states
and cannot be directly applied to $n>1$ equations from the 
Darboux covariant family (\ref{n-eq}): The non-abelian shape of, say, the $n=2$
equation 
\bea 
i\dot \rho
&=&
[A^2,\rho^2]+[A,\rho A\rho]
\eea
requires different tricks. 
The solutions discussed
in the previous subsection started with non-stationary $\rho$'s, since those
satisfying $[\rho,A]=0$
lead to the projector $P$ commuting with both $A$ and $\rho$, and the 
binary transformation is trivial. Still, there exists another 
class of stationary solutions of (\ref{n-eq}), obtained if $A\rho=-\rho A$. 
Now the projector $P$ will not, in general, commute with $\rho$ and
$A$, and the binary transformation may be nontrivial. 

Assume $[P,\rho]\neq 0$. 
For $n=2m$ one has $\sum_{k=0}^n (-1)^k=1$ and
\bea
i|\dot\varphi\rangle 
&=&
\Big(
A^n\rho-\mu A^{n+1}\Big)|\varphi\rangle=
z_\mu A^n|\varphi\rangle\label{15}
\eea
The formal solution of the latter equation is 
\bea
|\varphi(t)\rangle = e^{-iz_\mu A^n t}|\varphi(0)\rangle \label{sol even}.
\eea
For odd $n$ we get $\sum_{k=0}^n (-1)^k=0$ and 
\bea
i|\dot\varphi\rangle 
&=&
-\mu A^{n+1}|\varphi\rangle
\eea
implying 
\bea
|\varphi(t)\rangle = e^{i\mu A^{n+1} t}|\varphi(0)\rangle\label{sol odd} .
\eea
To have an `interesting' $\rho[1]$
we must make sure that the solution of the Lax pair 
is not an eigenstate 
of $A^{2m}$ (for $n=2m$) or $A^{2(m+1)}$ (for $n=2m+1$); otherwise we have a
problem similar to this with the eigenstates of $\Delta_a$. 
Let us note that $\rho$ which anticommutes with $A$ does commute with 
$A^2$ and therefore also with the generators of the time evolution 
given by (\ref{sol even}) and (\ref{sol odd}), and we can 
look for an eigenstate of $\rho - \mu A$ at $t=0$. 

For simplicity we shall check the trick again on the Euler-Arnold-von Neumann
equation. 
Let $\alpha_k$ and 
$\sigma_k$ be Dirac-$\alpha$ and Pauli matrices, respectively.  
Consider the Hamiltonian 
\bea
H=\alpha_1\otimes {\bf 1}_{2\times 2}+{\bf 1}_{4\times 4}\otimes \sigma_1,
\eea
and take a non-density-matrix stationary solution 
\bea
\xi =\alpha_2\otimes \sigma_2 + \alpha_3\otimes \sigma_3
\eea
satisfying $\xi H=-H\xi $. Set $\mu=i$ and take 
\bea
\langle\varphi|
&=&
(i,0,-1,0,-i,0,1,0)
\eea
satisfying $(\rho-iH)|\varphi\rangle=0$. Then
\bea
{}&{}&
\xi [1](t)=\nonumber\\
&{}&
\left(
\begin{array}{cccccccc}
{\frac{1 + {e^{4 t}}}{1 + {e^{8 t}}}} 
& 
{\frac{i {e^{4 t}}}{1 + {e^{8 t}}}} 
& 
0 
& 
{\frac{-1}{1 + {e^{8 t}}}} 
& 
{\frac{e^{8 t} - {e^{4 t}}}{1 + {e^{8 t}}}} 
& 
{\frac{-i {e^{4 t}}}{1 + {e^{8 t}}}} 
& 
0 
& 
{\frac{-e^{8 t}}{1 + {e^{8 t}}}}
\\  
{\frac{-i {e^{4 t}}}{1 + {e^{8 t}}}} 
& 
{\frac{{e^{4 t}}-1}{1 + {e^{8 t}}}} 
& 
{\frac{1}{1 + {e^{8 t}}}} 
& 
0 
& 
{\frac{i {e^{4 t}}}{1 + {e^{8 t}}}} 
& 
{\frac{-e^{8 t} - {e^{4 t}}}{1 + {e^{8 t}}}} 
& 
{\frac{{e^{8 t}}}{1 + {e^{8 t}}}} 
& 
0
\\  
0 
& 
{\frac{1}{1 + {e^{8 t}}}} 
& 
{\frac{-1 - {e^{4 t}}}{1 + {e^{8 t}}}} 
& 
{\frac{i {e^{4 t}}}{1 + {e^{8 t}}}} 
& 
0 
& {\frac{{e^{8 t}}}{1 + {e^{8 t}}}} 
& 
{\frac{{e^{4 t}}-e^{8 t}}{1 + {e^{8 t}}}} 
& 
{\frac{-i {e^{4 t}}}{1 + {e^{8 t}}}}
\\ 
{\frac{-1}{1 + {e^{8 t}}}} 
& 
0 
& 
{\frac{-i {e^{4 t}}}{1 + {e^{8 t}}}} 
& 
{\frac{1 - {e^{4 t}}}{1 + {e^{8 t}}}} 
& 
{\frac{-e^{8 t}}{1 + {e^{8 t}}}} 
& 
0 
& 
{\frac{i {e^{4 t}}}{1 + {e^{8 t}}}} 
& 
{\frac{e^{8 t} + {e^{4 t}}}{1 + {e^{8 t}}}}
\\  
{\frac{e^{8 t} - {e^{4 t}}}{1 + {e^{8 t}}}}
& 
{\frac{-i {e^{4 t}}}{1 + {e^{8 t}}}} 
& 
0 
& 
{\frac{-e^{8 t}}{1 + {e^{8 t}}}} 
& 
{\frac{1 + {e^{4 t}}}{1 + {e^{8 t}}}} 
& 
{\frac{i {e^{4 t}}}{1 + {e^{8 t}}}} 
& 
0 
& {\frac{-1}{1 + {e^{8 t}}}}
\\   
{\frac{i {e^{4 t}}}{1 + {e^{8 t}}}} 
& 
{\frac{-e^{8 t} - {e^{4 t}}}{1 + {e^{8 t}}}} 
& 
{\frac{{e^{8 t}}}{1 + {e^{8 t}}}} 
& 
0 
& 
{\frac{-i {e^{4 t}}}{1 + {e^{8 t}}}} 
& 
{\frac{{e^{4 t}}-1}{1 + {e^{8 t}}}} 
& 
{\frac{1}{1 + {e^{8 t}}}} 
& 
0
\\  
0 
& 
{\frac{{e^{8 t}}}{1 + {e^{8 t}}}} 
& 
{\frac{{e^{4 t}}-e^{8 t}}{1 + {e^{8 t}}}} 
& 
{\frac{-i{e^{4t}}}{1 + {e^{8t}}}} 
& 
0 
& 
{\frac{1}{1 + {e^{8t}}}} 
& 
{\frac{-1 - {e^{4t}}}{1 + {e^{8t}}}} 
& 
{\frac{i{e^{4t}}}{1 + {e^{8t}}}}
\\ 
{\frac{-e^{8 t}}{1 + {e^{8t}}}} 
& 
0 
& 
{\frac{i{e^{4t}}}{1 + {e^{8t}}}} 
& 
{\frac{e^{8 t} + {e^{4t}}}{1 + {e^{8t}}}} 
& 
{\frac{-1}{1 + {e^{8t}}}} 
& 
0 
& 
{\frac{-i{e^{4t}}}{1 + {e^{8t}}}} 
& 
{\frac{1 - {e^{4t}}}{1 + {e^{8t}}}}
\end{array}
\right).\nonumber
\eea
The eigenvalues of $\xi $ and $\xi [1](t)$ are $0$ and $\pm 2$. To produce the 
density matrix we shift the spectrum by $\Lambda\geq 2$ and 
rescale to get the unit trace.
This will be again a `self-scattering' solution  
but its explicit form will not be shown here. 
The second strategy has the advantage of being applicable to generically 
infinite-dimensional problems. A nontrivial question (requiring still more 
tricks) is how to produce trace-class solutions if the Hilbert space 
is not finite-dimensional.

\section*{Acknowledgments}

Our work was supported by the KBN Grant No. 2~P03B~163~15, the 
Flemish-Polish Project No. 007, and (MC)  Alexander von Humboldt-Stiftung.

\section*{Appendix}

Generic solutions of the equation of motion (\ref {EA}) with $H$
given by (\ref {H}) can be obtained by a straightforward integration. Change
the basis so that $H$ is of the form
\bea
H=-\left(\matrix{
\mu  &0    &0\cr
0    &-\mu &0\cr
0    &0    &\lambda\cr
}\right)
\label {3by3}
\eea
Then the diagonal elements of $\rho$ are constants of motion. Write down
the equation of motion for the remaining matrix elements.
One obtains
\bea
i\dot\rho_{1,2}&=&2\mu 
[(\rho_{1,1}+\rho_{2,2})\rho_{1,2}
+\rho_{1,3}\overline{\rho_{2,3}}],\cr
i\dot\rho_{1,3}&=&(\mu-\lambda)
[(\rho_{1,1}+\rho_{3,3})\rho_{1,3}
+\rho_{1,2}\rho_{2,3}],\cr
i\dot\rho_{2,3}&=&-(\mu+\lambda)
[(\rho_{2,2}+\rho_{3,3})\rho_{2,3}
+\overline{\rho_{1,2}}\rho_{1,3}].
\label {snled}
\eea
Let $W=|\rho_{1,2}|^2$. It is not difficult to show that
it satisfies the equation
\bea
\ddot W=aW^2+bW+c
\label {cleq}
\eea
with $a$, $b$, and $c$ constants.
Equation (\ref {cleq}) is well known \cite {DA}.
Its solutions are doubly-periodic functions.
The result is of the form
\bea
W(t)=\beta^{-1}{\rm sn}^2(\alpha(t-t_0),k)+\gamma
\eea
with sn the Jacobi elliptic function, and
with $k$, $\alpha$, $\beta$, $\gamma$, and $t_0$ constants.
In the limit $k=1$ one of the periods becomes infinite.
The result (\ref {finsol}) obtained by the binary Darboux
transformation corresponds
precisely to this $k=1$ solution. It is at the moment not clear to us
what kind of a seed solution (if any) can lead, via the binary Darboux 
transformation, to the $k\neq 1$ class. An algebraic characterization 
of such additional seed solutions
would be important from the perspective of more general 
cases, especially the infinite-dimensional ones.

\end{document}